\documentclass[letterpaper,12pt]{article}
\usepackage{graphicx}
\usepackage{color,amssymb,enumerate,float,amsmath}
\usepackage{tikz}
\usetikzlibrary{matrix}
\usepackage{caption}
\usepackage{subcaption}
\usepackage{cite}
\usepackage{ifpdf}
\ifpdf
	\usepackage[pdftex,unicode,implicit]{hyperref}

	\hypersetup{
  	pdftitle     = {}, 
  	pdfkeywords  = {},
  	pdfauthor    = {},
  	pdfcreator   = {pdf\LaTeXe\ with package \flqq hyperref\frqq},
  	pdfproducer  = {pdf\LaTeXe\ with package \flqq hyperref\frqq},
  	pdfpagemode  = UseNone,  
  	pdffitwindow = true,  
  	unicode      = true,
  	plainpages   = true,
  	colorlinks   = true,  
  	citecolor    = black,  
  	urlcolor     = blue, 
  	linkcolor    = black
	}

\else

  \usepackage[dvips]{graphicx}
  \usepackage[unicode,implicit]{hyperref}

\fi 

\makeatletter
\@addtoreset{equation}{section}
\makeatother

\begin{document}
	
\thispagestyle{empty}

\begin{center}
{\bf \LARGE Casimir effect of rough plates under a magnetic field in Ho\v rava-Lifshitz theory }
\vspace*{15mm}

{\large Byron Droguett}$^{1,a}$
{\large and Claudio B\'orquez}$^{2,b}$
\vspace{3ex}

$^1${\it Department of Physics, Universidad de Antofagasta, 1240000 Antofagasta, Chile.}

$^2${\it Facultad de Ingenier\'ia, Arquitectura y Diseño, Universidad San Sebasti\'an, Lago Panguipulli 1390, Puerto Montt, Chile.
}
\vspace{3ex}

$^a${\tt byron.droguett@uantof.cl},
$^b${\tt claudio.borquez@uss.cl}

\hspace{.5em}

{\bf Abstract}
\begin{quotation}{\small\noindent
} 
We investigate the Casimir effect for parallel plates within the framework of Ho\v rava-Lifshitz theory in $3+1$ dimensions, considering the effects of roughness, anisotropic scaling factor, and an uniform constant magnetic field. Quantum fluctuations are induced by an anisotropic charged-scalar quantum field subject to Dirichlet boundary conditions. To incorporate surface roughness, we apply a coordinate transformation to flatten the plates, treating the remaining roughness terms as potential. The spectrum is derived using perturbation theory and regularized with the $\zeta$-function method. As an illustrative example, we consider plates with periodic boundary conditions.
\end{quotation}
\vspace{3ex}
\end{center}

\thispagestyle{empty}

\newpage


\section{Introducction}


The Casimir effect is a phenomenon predicted by quantum field theory arising from vacuum fluctuations at the quantum level. H. B. Casimir demonstrated that two parallel, uncharged, and isolated plates, separated by a distance much smaller than their length, experience an attractive force due to the quantum fluctuations of the electromagnetic field under specific boundary conditions \cite{Casimir:1948dh}. This effect has been experimentally confirmed with high precision, making it an ideal test to study the properties of quantum vacuum fluctuations \cite{Lamoreaux:1996wh, Bressi:2002fr}. The effect has been observed in different geometries, and in some cases the force is repulsive, indicating that the Casimir effect is influenced by the geometry of the system \cite{Boyer:1968uf}. Various factors have been shown to influence the Casimir effect. Research indicates that boundary conditions related to specific materials, spacetime topology, magnetic field, and temperature also modified the energy spectrum \cite{Bordag:2001qi, Teo:2011kt, Zhao:2006rr,Beneventano:2004zd,Beneventano:2005sd}. The dimensionality of spacetime is a key factor in the Casimir effect, and research on two-dimensional materials, such as the graphene family, has become crucial for technological advances in material science \cite{Bellucci:2019ybj}.

We aim to investigate the Casimir effect in theories characterized by the breaking of Lorentz symmetry, particularly within Ho\v rava-Lifshitz-like frameworks. These theories are based on the establishment of an anisotropy between space and time through an anisotropic scaling factor \cite{Anselmi:2008bq}. Several cases exhibiting anisotropic behavior have been analyzed, including extensions involving Klein-Gordon and fermionic fields \cite{ Ferrari:2010dj, MoralesUlion:2015tve, daSilva:2019iwn, deMello:2024lmi}. Other research on Lorentz violation has introduced terms in the Lagrangian that incorporate a preferred direction \cite{Cruz:2017kfo,deMello:2022tuv}. The effects of the magnetic field and finite-temperature have also been studied \cite{Erdas:2013jga,Erdas:2015yac,Erdas:2023wzy,Erdas:2020ilo,Cruz:2018bqt,Erdas:2021xvv, Cheng:2022mwd}.
In theories of gravity, Ho\v rava-Lifshitz gravity emerges as a promising candidate to complement the high-energy regime of general relativity \cite{Horava:2009uw, Blas:2009qj}. The established anisotropy reduces the general diffeomorphism invariance of the theory and breaks the Lorentz symmetry in the ultraviolet regime. The reduction of symmetry leads to the emergence of an instantaneous scalar mode. The foliation of the theory is defined as a set of spatial hypersurfaces accompanied by a preferred temporal direction, which must be preserved by reduced diffeomorphisms. The most
appropriate variables to describe this preferred foliation are the ADM variables \cite{Arnowitt:1962hi}. Recently, the renormalization and unitarity of the nonprojectable Ho\v rava theory has been demonstrated, making it a consistent quantum gravity theory \cite{Bellorin:2023dwk,Bellorin:2024qyy}.
In previous work \cite{Borquez:2023cuf}, we computed the Casimir effect of an anisotropic scalar field on a membrane embedded in a conical $2+1$-dimensional manifold, induced by the presence of a massive located point-particle at rest. Consequently, we found that the Casimir energy and force depend on the presence of the massive particle, the anisotropic scaling factor, and the temperature.

In this research, we analyze the modifications to the Casimir energy spectrum of an anisotropic charged-scalar field, caused by the presence of roughness on conducting, uncharged parallel plates embedded in a $3+1$-dimensional spacetime manifold. Additionally, we consider the introduction of an external constant uniform magnetic field inside the parallel plates and oriented perpendicular to them, where the contributions of the weak and strong field limits are calculated. For a more realistic application, we treat the roughness as a perturbation of the flat case. By performing an appropriate change of coordinates, the parallel plates are now a flat surface and the remaining terms associated with the roughness are incorporated into the potential term \cite{Borquez:2023ajx, Droguett:2014gqa,Droguett:2024tpe}. To obtain the spectrum of eigenvalues, we apply perturbation theory and use the $\zeta$--function for the regularization process \cite{Kirsten:2010zp}. Finally, we provide a specific example of plates with periodic behavior. 

This paper is organized as follows. In Sect. 2, we present the problem of the rough plates considering a constant and uniform magnetic field. In Sect. 3, we apply the regularization method using the $\zeta$--function and determine the energy and force density. We present the contribution of the weak and strong magnetic field and an explicit example of plates with periodic border. In Sect. 4, we present our conclusions.


\section{Ho\v rava-Lifshitz-like charged-scalar field}


\subsection{The anisotropic operator}

The Ho\v rava-Lifshitz gravity theory \cite{Horava:2009uw,Blas:2009qj} is formulated be invariant under anisotropic scaling between space and time, expressed by
\begin{equation}
	[t]=-z\,,
    \qquad
    [x^i]=-1 \,,
\label{scalespacetime}
\end{equation}
where $z$ represents the anisotropy scaling factor. This leads to a reduction in the symmetry of general diffeomorphisms and a breaking of Lorentz symmetry in the ultraviolet regime, making this theory a candidate to complete the high-energy regime of general relativity. Consequently, the symmetry that preserves the foliation  takes the form $\delta t = \zeta^0(t), x^i = \zeta^i(t,x^k)$, where the time is reparameterized on itself, giving the foliation an absolute physical meaning. In the context of gravity, the most suitable variables to describe this foliation are the ADM variables. It is possible to generalize field theories such as the Klein–Gordon, quantum electrodynamic, and Yang–Mills theories, where the quantum fluctuations are generated by the anisotropic field. In particular, we investigate the Casimir effect arising from the quantum fluctuations of an anisotropic charged-scalar quantum field in the presence of rough, conducting, parallel, uncharged plates embedded in a $3+1$ dimensional spacetime manifold. The analysis assumes Dirichlet boundary conditions and considers an external constant uniform magnetic field perpendicular to the plates. Therefore, the generalization of the Lagrangian density associated with an anisotropic charged-scalar field is given by
\begin{eqnarray}
    \mathcal{L} =
    N\sqrt{g}
    \left(
    \partial_t\Phi^*\partial_t\Phi
    - l^{2(z-1)}D^*_{i1}\cdots D^*_{iz}\Phi^* D^{i1}\cdots D^{iz}\Phi
    \right)
    \,,
    \label{action}
\end{eqnarray}
with $l\approx 1/\Lambda_{l}$, being $\Lambda_{l}$ the energy scale introduced to keep the physical dimensionality correct\footnote{In this work, we adopt natural units in which \( \hbar = c = 1 \).} {\cite{Anselmi:2008bq}. The extended covariant spatial derivative has the form $D_j = \nabla_j + iqA_j$, where $q$ is the charge of the field, $A_j$ is the vector potential field and $\nabla_j$ is the covariant derivative associated to spatial metric. For consistency, in $d=3$ dimensions, the vector $A_j$ and charge scale as follow
\begin{equation}
    [A_{i}] = \frac{3-z}{2}
    \,,
    \qquad
    [q] = \frac{z-3}{2} + 1
    \,.
\end{equation}
The equation of motion for the charged-scalar field is
\begin{equation}
\label{EKGeq}
    \left(\partial_t^2+(-1)^{z}l^{2(z-1)}\left(D^{2}\right)^{z}\right)\Phi
    = 0\,,
\end{equation}
and the spatial operator is defined by
\begin{eqnarray}
    D^{2}
    =
    \Delta
    + 2iqA_j\nabla^{j} 
    + iq\nabla^j A_j
    - q^2g^{jk}A_j A_k
    \,.
    \label{DD}
\end{eqnarray}
Gauge invariance allows us to choose the form of the vector potential. Specifically, we set $A_i=\left(-By,0,0\right)$.

On the other hand, the geometry of the parallel conducting plates is modeled by the Cartesian coordinates, where the $w$ coordinate is bordered by $0\leq w\leq a + f(x,y)$, with $(x,y)\in\mathbb{R}^{2}$. The parameter  $a$ represents its width and $f(x,y)$ contains all the information about the roughness of the plates, with the assumption that $f(x,y)\ll a$. Taking these conditions into account, we implement a change of variable on the $w$ coordinate, in such a way that the plates exhibits flat borders. Then, we define  $w = \rho\left( 1 + f(x,y)/a\right)$, with  $0\leq\rho\leq a$. The metric associated to these new coordinates is represented by\footnote{It is possible to configure perturbations at both plates so that the contributions can be expressed as a linear combination of the perturbations, hence these can be additive or cancel. In this sense, the conclusions about how the magnetic field and perturbation terms affect the Casimir force remain consistent.} 
\begin{eqnarray}
    g_{ij} &=& 
    \left(
    \begin{array}{ccc}
    1 + 
    \left(
    \rho/a
    \right)^2(\partial_x f)^2
    & \left(
    \rho/a
    \right)^2 \partial_x f \partial_yf 
    & \frac{\rho}{a} \left(1+\frac{f}{a}\right) \partial_xf
    \\
    \left(
    \rho/a
    \right)^2 \partial_x f \partial_y f  
    & 1 + 
    \left(
    \rho/a
    \right)^2
    (\partial_yf)^2 
    & \frac{\rho}{a} 
    \left(
    1+\frac{f}{a}
    \right) \partial_yf
    \\
    \frac{\rho}{a}
    \left(
    1+\frac{f}{a}
    \right) \partial_xf
    & \frac{\rho}{a}  
    \left(
    1+\frac{f}{a}
    \right) \partial_yf
    & \left(
    1+\frac{f}{a}
    \right)^2 
    \\
    \end{array}
    \right)
    \,.
    \label{metric}
\end{eqnarray}
The Laplace Beltrami operator associated to the metric (\ref{metric}) on the scalar field, is given by
\begin{eqnarray}
    \Delta\Phi
    &=&
    \Delta_x\Phi
    + \Delta_y\Phi
    + \frac{1}{(a+f)^2}\left[
    \rho^{2}\left((\partial_{x}f)^{2} + (\partial_{y}f)^{2}\right)
    + a^2\right]\Delta_{\rho}\Phi
    \nonumber
    \\&&
    + \frac{\rho}{(a+f)^2}\left[
    2\left((\partial_{x}f)^{2} + (\partial_{y}f)^{2}\right)
    - (a+f)\left(\Delta_{x}f + \Delta_{y}f\right)
    \right]\partial_{\rho}\Phi
    \nonumber
    \\&&
    - \frac{2\rho}{a+f}\left(
    \partial_{x}f\Delta_{x\rho}\Phi
     + \partial_{y}f\Delta_{y\rho}\Phi\right)
     \,.
\end{eqnarray}
Conveniently, we perform the following change of variables to obtain dimensionless coordinates:
\begin{eqnarray}
    \begin{split}
    x & = u_1 L_1\,, \qquad -1\leq u_1\leq 1\,,
    \\
    y & = u_2 L_2\,, \qquad  -1\leq u_2\leq 1\,,
    \\
    \rho & = va \,,\qquad  \qquad 0\leq v\leq 1\,.
    \end{split}
    \label{finalchangeofvariable}
\end{eqnarray}
The perturbative nature of the function $f$ (from now on $\hat{f}=f(u_1L_1,u_2L_2)$ in the new coordinates) also allows us to define that $L_{1,2}\gg (\hat{f},\partial_{i} \hat{f},\partial^{2}\hat{f})$, leading to a helpful simplification of the Laplace-Beltrami operator. Because both parameters $L_{1}$ and $L_{2}$ are infinitely large compared to the distance $a$ and the perturbation $\hat{f}$, we simply set $L_{1}=L_{2}=L$. For the remaining terms, if we consider perturbations of $\hat{f}$ up to second order, we have
\begin{eqnarray}
    \Delta\Phi
    =
    \left(\frac{1}{L^2}\partial^2_{u_1}
    + \frac{1}{L^2}\partial^2_{u_2}
    + \frac{1}{a^2}\partial^2_v
    - \mathcal{M}(u_1,u_2)\partial_v^2\right)\Phi\,,
\label{Delta}
\end{eqnarray}
where $ \mathcal{M}$ is defined by
\begin{eqnarray}
    \mathcal{M}(u_1,u_2) = 
    \frac{2\hat{f}}{a^{3}}
    - \frac{3\hat{f}^{2}}{a^4}
    \,.
    \label{M}
\end{eqnarray}
Finally, considering the operators (\ref{DD}) and (\ref{Delta}), and taking into account the gauge-fixing mentioned above, we get the total operator associated with the eigenvalue  problem
\begin{eqnarray}
    -\Delta\Phi
    + 2iqBu_2\partial_{u_1}\Phi
    + \left(qBLu_2\right)^2\Phi
    =\lambda\Phi
    \,.
\end{eqnarray}
The Dirichlet boundary conditions are:
\begin{eqnarray}
    \Phi(u_{1},u_{2},0) 
    =
    \Phi(u_{1},u_{2},1)
    = 0
    \,,
\label{DBC}
\end{eqnarray}
therefore, the total spectrum associated with an anisotropic operator $\mathcal{P}$ satisfying these boundary conditions, is given by
\begin{eqnarray}
    \mathcal{P}\Phi 
    =
    (-1)^{z}l^{2(z-1)}D^{2z}\Phi
    =
    l^{2(z-1)}\lambda^z\Phi\,.
\end{eqnarray}


\subsection{The perturbation theory}

As a first step, we consider the zeroth-order case in perturbations, assuming that $z=1$, 
\begin{eqnarray}
    \left(
    - \frac{1}{L^2}\partial^2_{u_1}
    - \frac{1}{L^2}\partial^2_{u_2}
    - \frac{1}{a^2}\partial_v^2
    + 2iqBu_2\partial_{u_1}
    + \left(qBLu_2\right)^2\right)
    \Phi^{(0)} = 
    \lambda^{(0)}\Phi^{(0)}
    \,.
    \label{Eqz=1}
\end{eqnarray}
After applying a coordinate transformation to (\ref{Eqz=1}), we can arrive at the well-known Schrödinger equation for a harmonic oscillator. Then, the normalized eigenfunctions\footnote{The normalization is achieved by considering $L$ tending to infinity.} satisfying boundary conditions (\ref{DBC}) can be written in terms of Hermite’s polynomials  as follows,
\begin{eqnarray}
    \Phi^{(0)}_{n,m}(k)
    = 
    \left(\frac{\sqrt[4]{\alpha}}{n!2^{n}\sqrt{\pi}}\right)^{1/2}
    e^{-\frac{\sqrt{\alpha}}{2} \left( u_2-  \beta\right)^2}
    H_{n}\left(\sqrt[4]{\alpha } \left( u_2 - \beta \right)\right)
    e^{iku_1}
    \sin\left(m\pi v\right)
    \,,
\end{eqnarray}
where $\alpha = (qBL^2)^2$, $\beta = k/qBL^2$ and $k$ is the eigenvalue of the $u_{1}$ coordinate. The corresponding zeroth-order eingenvalue is
\begin{equation}
    \lambda^{(0)}_{n,m} =
    (2n + 1)qB 
    + \left( \frac{m\pi}{a}\right)^2\,, 
    \qquad n\in \mathbb{N}_0\,, m\in \mathbb{N}
    \,.
    \label{EigenLambda}
\end{equation}

To find the first-order eigenvalues in perturbation theory, we must calculate the following integral
\begin{eqnarray}
    \lambda_{n,m}^{(1)} =
    \int_{-1}^1\int_{-1}^1\int_0^1\,du_1\,du_2\,dv\,
    \Phi^{*(0)}_{n,m}(k)\,\mathcal{M}(u_1,u_2)\,\partial^2_v\,\Phi_{n,m}^{(0)}(k).
\end{eqnarray}
The integration over $v$ can be performed straightforwardly; however, the integrals over the coordinates $u_{1}$ and $u_{2}$ become significantly more laborious if $\mathcal{M}$ depends on both $u_{1}$ and $u_{2}$. After integrating out $v$, we obtain
\begin{eqnarray}
    \lambda_{n,m}^{(1)} =
    - \frac{(m\pi)^2}{2}
    \left(
    \frac{\sqrt[4]{\alpha}}{n!2^{n}\sqrt{\pi}}
    \right)
    \int_{-1}^1\int_{-1}^1\,du_1\,du_2\,
    e^{-\sqrt{\alpha}
    \left( 
    u_{2}
    - \beta
    \right)^2} 
    H^2_{n}
    \left(
    \sqrt[4]{\alpha }
    \left(
    u_2 - \beta
    \right)
    \right)
    \mathcal{M}(u_1,u_2).
    \nonumber\\
\end{eqnarray}
To simplify the calculations, we assume that the geometric perturbation $\mathcal{M}$ depends only on the coordinate $u_{1}$. Under this assumption, the integral over $u_{2}$ can be performed more efficiently. Furthermore, although our method is applicable to any type of roughness, from this point forward, we will restrict our analysis to periodic functions. Then, the total eingenvalue\footnote{The analysis of perturbation theory is highly simplified by assuming that the perturbative roughness function is periodic. Consequently, from the \emph{second-order of eigenvalue perturbations} onward, there are no contributions to the second-order perturbations of the roughness.} is
\begin{eqnarray}
    \lambda_{n,m}
    &=&
    (2n+1)qB
    + \left(
    m\pi
    \right)^2
    \left[
    \frac{1}{a^2}
    - \frac{1}{2}
    \int_{-1}^1\,du_1
    \mathcal{M}(u_1)
    \right].
\end{eqnarray}


\section{ $\zeta$–function regularization}

We address the regularization of the spectrum associated to spatial operator using the $\zeta$--function method, which is given by
\begin{eqnarray}
    \zeta_{\mathcal{P}}(s)
    =
    l^{-2(z-1)s}\frac{qBL^2}{2\pi}
    \sum_{\substack{n=0 \\ m=1}}^{\infty}
    \left[
    (2n+1)qB
    + \left(
    m\pi\right)^2
    \left(
    \frac{1}{a^2}
    - \frac{1}{2}
    \int_{-1}^1\,du_1
    \mathcal{M}(u_1)
    \right)
    \right]^{-zs},
    \label{zetaP}
\end{eqnarray}
where the factor $qBL^2/2\pi$ has been taken into account due to the degeneracy of the Landau levels. This function has the structure of the Epstein $\zeta$–function, hence the integral take the form
\begin{eqnarray}
    \zeta_{\mathcal{P}}(s)
    =
    l^{-2(z-1)s}\frac{qBL^2}{2\pi\Gamma(zs)}
    \int_0^\infty\,dt\, t^{zs-1}
    \sum_{\substack{n=0 \\ m=1}}^{\infty}
    \exp
    \left[
    -t\left(
    (2n+1)qB + rm^2
    \right)
    \right],
\label{zetaPgeneral}
\end{eqnarray}
where 
\begin{eqnarray}
    r
    =
    \pi^2
    \left(\frac{1}{a^2}
    -\frac{1}{2}
    \int_{-1}^1\,du_1
    \mathcal{M}(u_1)
    \right)\,.
    \label{rreduce}
\end{eqnarray}


\subsection{Weak magnetic field limit}

We consider the weak magnetic field case for fixed $m$ when $qBa^2\ll1$. In Eq. (\ref{zetaPgeneral}) the summation over $n$ can be expressed by
\begin{eqnarray}
    \sum_{n=0}^\infty \exp
    \left[
    -t((2n+1)qB)
    \right]
    =
    \frac{1}{2\sinh{(qBt)}}
    \,,
    \label{zetaPhyperbolic}
\end{eqnarray}
thus, the $\zeta$--function can be written as 
\begin{eqnarray}
    \zeta_{\mathcal{P}}(s)
    =
    l^{-2(z-1)s}\frac{qBL^2}{4\pi\Gamma(zs)}
    \int_0^\infty\,dt\, \frac{t^{zs-1}}{\sinh{(qBt)}}
    \sum_{m=1}^\infty\exp
    \left(-rm^2t
    \right)
    \,.
    \label{zetasinh}
\end{eqnarray}
At this point, we can do a series expansion over $qB$ up to second order as follows,
\begin{eqnarray}
    \frac{qB}{\sinh{(qBt)}}
    = 
    \frac{1}{t} 
    - \frac{q^2B^{2}}{6}t 
    \,.
\end{eqnarray}
With this approximation, the integration in Eq. (\ref{zetasinh}) can be performed directly. Considering the summation over $m$, the result is expressed in terms of the Riemann $\zeta$--function,
\begin{eqnarray}
    \zeta_{\mathcal{P}}(s)
    =
    \frac{l^{-2(z-1)s}L^2}{4\pi\Gamma(zs)}
    \left(
    {r}^{1-zs}\Gamma
    \left(
    zs - 1
    \right)\zeta_{R}(2zs-2)
    - \frac{(qB)^{2}}{6\,r^{1+zs}}\Gamma
    \left( 
    zs+1 
    \right)\zeta_{R}(2zs+2)
    \right)\,.
    \nonumber\\
    \label{zetaintegrated}
\end{eqnarray}
To determine the Casimir energy, we must evaluate $s=-1/2$ and focus on the finite components of the $\zeta$--function for different values of $z$. Specifically in the case $z=1$, divergent terms arise from the $\zeta_{R}$--function in the second term of equation (\ref{zetaintegrated}). To avoid these divergences, we perform a series expansion around $s=-1/2$. 
Then, the Casimir energy density for $z=1$ is\footnote{The charged scalar field correspond to $\Phi=\frac{1}{\sqrt{2}}(\phi_{1}+i\phi_{2})$, where $\phi_{1},\phi_{2}$ are real scalar fields. This means that the system has two degrees of freedom, hence the Casimir energy density is $\mathcal{E}_{C}=\zeta(-1/2)$}.
\begin{eqnarray}
    \mathcal{E}_{C}^{z=1} =
    -\frac{r^{3/2}}{720\,\pi }
    + \frac{r^{-1/2}q^2B^{2}}{48\,\pi}\left( \gamma
    - 1 - \frac{1}{2}\ln(r) \right)\,.
\end{eqnarray}
In the $z\neq 1$ case, the Casimir energy can be expressed in the general form
\begin{eqnarray}
    \mathcal{E}_{C}(z)
    &=&
    \frac{l^{z-1}}{24\pi\Gamma(-z/2)}
    \left[
    6\,r^{1 + z/2}\zeta_{R}(-z-2)\Gamma
    \left(
    - 1 - \frac{z}{2}
    \right)
    \right.
    \nonumber
    \\&&
    \left.
    - r^{-1+z/2}(qB)^{2}\zeta_{R}( 2 - z )\Gamma
    \left( 
    1 - \frac{z}{2}
    \right)
    \right]
    \,.
\label{casimirdensityenergyweak}
\end{eqnarray}
The force density is obtained by differentiating with respect to the separation $a$ between the plates. Considering both the surface roughness described in Eq. (\ref{rreduce}) and the weak magnetic field up to second-order perturbations, the Casimir force density for \( z = 1 \) and \( z \neq 1 \) is given, respectively, by:
\begin{eqnarray}
    \mathcal{F}^{z=1}_{C}
    &=&
    - \frac{\pi^{2}}{240\,a^{4}}
    + \frac{q^{2}B^{2}}{48\,\pi^{2}}
    \left( 
    \ln
    \left(
    \frac {\pi }{a}
    \right) 
    - \gamma
    \right)
    + \frac{\pi^2}{120\,a^{5}}\int_{-1}^{1}\hat{f}du_{1}
    \nonumber\\
    &&
    - \frac{\pi^{2}}{192\,a^{6}}\left[\frac{1}{2}\left(\int_{-1}^{1}\hat{f}du_{1}\right)^{2}
    + 3\int_{-1}^{1}\hat{f}^{2}du_{1}\right]
    \,,
    \label{forcedensityfinalz=1}
\end{eqnarray}
\begin{eqnarray}
    \mathcal{F}_{C}(z)
    &=&
    \frac{l^{z-1}}{24\pi\Gamma(-z/2)}
    \left\{
    - 3\pi^{z+2}(z+2)\zeta_{R}(-z-2)\Gamma\left( 
    - 1 - \frac{z}{2}
    \right)
    \left[
    - \frac{2}{a^{z+3}}
    \right.
    \right.
    \nonumber
    \\
    &&
    + \frac{(z+3)}{a^{z+4}}\int_{-1}^1\hat{f}\,du_1
    - \frac{(z+4)}{4\,a^{z+5}}
    \left(
    z\left(
    \int_{-1}^1\hat{f}\,du_1
    \right)^2
    \right.
    \nonumber
    \\
    &&
    \left.
    \left.
    + 6\int_{-1}^1\hat{f}^{2}\,du_1
    \right)
    \right]
    \left.
    - (z-2)\pi^{z-2}\zeta_{R}( 2 - z )\Gamma\left( 1 - \frac{z}{2}\right)
    \frac{(qB)^{2}}{a^{z-1}}
    \right\}\,.
    \label{forcedensityfinal}
\end{eqnarray}
We observe that when the anisotropic scaling factor $z$ takes an even value, the $\zeta$--function approaches zero, causing both the Casimir energy (\ref{casimirdensityenergyweak}) and force (\ref{forcedensityfinal}) to vanish for these values. To illustrate these results, we proceed to calculate the Casimir effect for a specific case considering periodic roughness. We choose the following periodic function in the original coordinates $f(x)=\xi\cos(2\pi x/L)$, where $\xi$ represents a small parameter that indicates the perturbative nature of $f$. This choice of $f$ allows for further simplification of Eqs. (\ref{forcedensityfinalz=1}) and (\ref{forcedensityfinal}). Therefore, the force densities take the form\footnote{The terms in the force density that are independent of $a$ originate from the uniform energy density, meaning they are proportional to the volume between the plates $L^2a$. If we consider the presence of a magnetic field both outside and inside, these terms do not contribute to the force density \cite{Erdas:2013jga}. In this work, we consider the presence of a magnetic field inside the plates, and it will be applied both in the weak and strong field limit (see Sec. \ref{strong}).}
\begin{eqnarray}
    \mathcal{F}_{C}^{z=1}
    =
    - \frac{\pi^{2}}{240\,a^{4}}
    - \frac{\pi^{2}\xi^{2}}{64\,a^{6}}
    + \frac{q^{2}B^{2}}{48\,\pi^{2}}
    \left( 
    \ln
    \left(
    \frac {\pi }{a}
    \right) 
    - \gamma
    \right)\,,
\end{eqnarray}
\begin{eqnarray}
    \mathcal{F}_{C}(z)
    &=&
    \frac{l^{z-1}}{24\pi\Gamma(-z/2)}
    \left\{
    - 3\pi^{z+2}(z+2)\zeta_{R}(-z-2)\Gamma
    \left( 
    - 1 - \frac{z}{2}\right)
    \left[
    - \frac{2}{a^{z+3}}
    \right.
    \right.
    \nonumber
    \\
    &&
    \left.
    \left.- \frac{3(z+4)}{2a^{z+5}}\xi^2
    \right]
    - (z-2)\pi^{z-2}\zeta_{R}( 2 - z )\Gamma
    \left( 
    1 - \frac{z}{2}
    \right)
    \frac{{q}^{2}{B}^{2}}{a^{z-1}}
    \right\}\,.
    \label{FuerzaZ}
\end{eqnarray}

In Fig. (\ref{Fig1}), the behavior of the force density as a function of the separation distance is shown for various values of the anisotropic factor $z$. We observe that the magnitude of the force density increases in each case due to additional contributions from the second-order perturbation in $\xi$, regardless of the magnetic field. As noted above, for cases where $z$ takes even values, the force density is zero. When considering the effect of a weak magnetic field (while neglecting the roughness factor), a contribution to the second-order perturbation is also observed. However, this contribution remains very close to the planar case (see subfigure in Fig. \ref{Fig1}). In general, the presence of the weak magnetic field minimally reduces the Casimir force, as illustrated in the subfigure for $z=1$. Unlike in the case of $z = 1$, the force density in (\ref{FuerzaZ}) depends on the scaling factor $l$ for any value of the anisotropic factor. As higher values of $z$ are chosen, the force density strongly depends on $l$ decreasing in magnitude, as shown in Fig. \ref{Fig1}. The anisotropic scaling factor plays a crucial role in determining the direction of the force. For example, in the case of $z =3$, the force becomes repulsive. On the other hand, when taking into account the roughness we notice that in each case the force density increases in magnitude compared to the planar case. This can be interpreted as the presence of roughness makes the force more detectable in measurements. 

\begin{figure}[H]
    \begin{minipage}[b]{0.8\linewidth}
    \centering
\includegraphics[width=.8\linewidth]{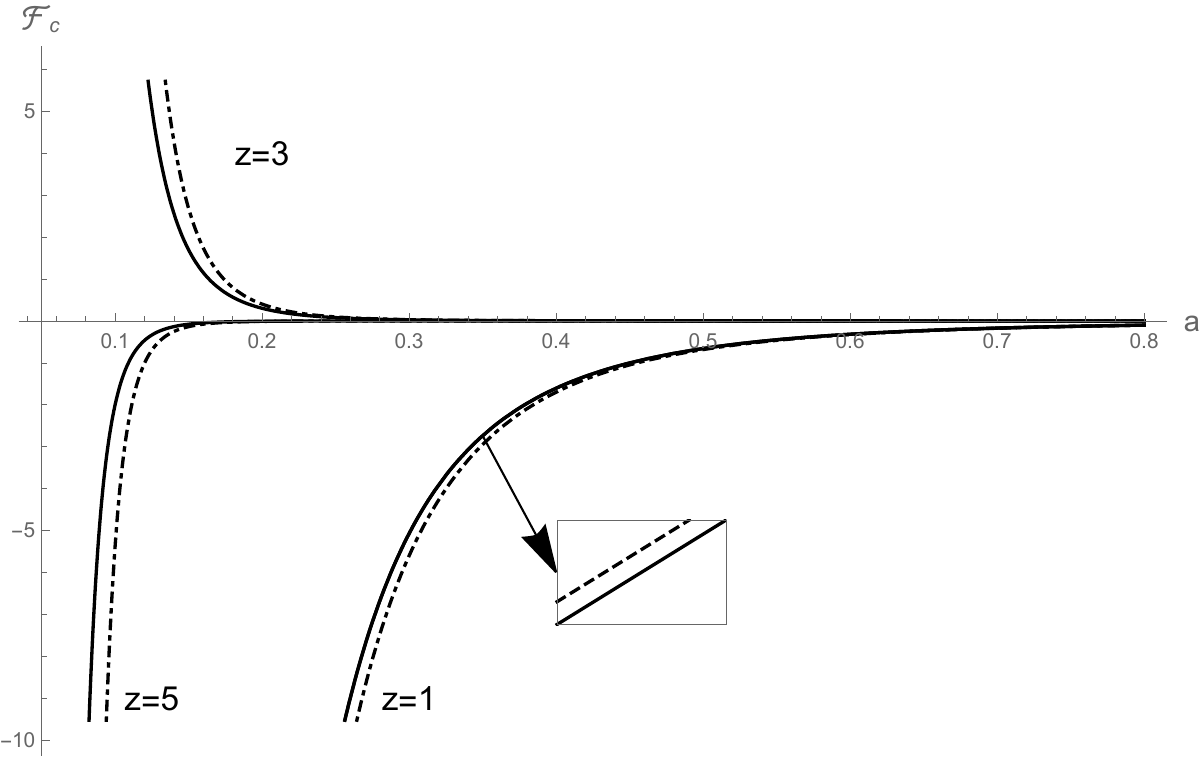} 
    \vspace{4ex}
  \end{minipage}
  \caption{Casimir force density as a function of the separation distance $a$.  We present the curves for different values of the anisotropic scaling factor: $z=1,3,5$. For the cases $z=3,5$, we set $l=0,01$. The solid curves represent the case without perturbative terms, while the dashed-point curves represent the case with surface roughness $\xi=0.05$ (units of lenght) and no magnetic field. The subfigure shows a close-up of the solid curve in the range $a\sim 10^{-6}$ (units of length) and $\mathcal{F}_{C}\sim 10^{-5}$ (units of lenght$^{-4}$), where it is observed that the dashed curve reflects the small contribution from the magnetic field $qB=0.1$ (units of lenght$^{-2}$)}.
  \label{Fig1}
\end{figure}


\subsection{Strong magnetic field limit}\label{strong}

We consider the strong magnetic limit $qBa^2\gg 1$ for fixed $m$. In this case, from Eq. (\ref{zetaPgeneral}), we take the following asymptotic behavior
\begin{eqnarray}
    \frac{qB}{\sinh{(qBt)}} 
    \simeq
    2qBe^{-qBt} 
    \,,
\end{eqnarray}
and we apply the Poisson resumation on $m$. Then, the $\zeta$--function is expressed as
\begin{eqnarray}
    \zeta_{\mathcal{P}}(s)
    =
    l^{-2(z-1)s}\frac{qBL^2}{4\pi\Gamma(zs)}
    \int_0^\infty\,dt\, t^{zs-1}e^{-qBt}
    \left[
    - 1
    + \sqrt{\frac{\pi}{rt}}
    + 2\sqrt{\frac{\pi}{rt}}
    \sum_{m=1}^\infty\exp
    \left(
    -\frac{\pi^2m^2}{rt}
    \right)
    \right].
    \nonumber\\
\end{eqnarray}
Here we observe that the term containing the summation over $m$ decays rapidly for $m > 1$, even when taking into account the presence of the perturbative factor $r$. After performing the integrations, we obtain
\begin{eqnarray}
    \zeta_{\mathcal{P}}(s)
    &=&
    \frac{l^{-2(z-1)s}L^2}{4\pi}
    \left[
    - (qB)^{-zs+1}
    + \frac {r^{-1/2}\,(qB)^{-zs+3/2}\,\pi^{1/2}\,\Gamma
    \left(
    zs - \frac{1}{2}
    \right)}{\Gamma(zs)}
    \right.
    \nonumber
    \\&&
    \left.+ \frac {4\,(qB)^{\frac{1}{2}(-zs + 5/2)}\,\pi^{zs}\,r^{-\frac{1}{2}(zs + 1/2)}}{\Gamma(zs) }K_{zs-1/2}
    \left(
    2\pi\sqrt{\frac{qB}{r}}
    \right)
    \right]
    \,,
    \label{zetaZs}
\end{eqnarray}
where $K_{zs-1/2}$ is the modified Bessel function of second kind.
Due to the strong-magnetic field, we can take the asymptotic limit of the modified Bessel function $K_{\nu}(\sigma)\sim\sqrt{\frac{\pi}{2\sigma}}e^{-\sigma}$, making the last term in the above equation more handle. To determine the Casimir energy, we must evaluate $s=-1/2$ and focus on the finite components of the $\zeta$--function for different values of $z$. For any $z$ value, divergent terms arise from the $\Gamma$--function in Eq. (\ref{zetaZs}). To avoid these divergences, we perform a series expansion around $s=-1/2$.  Then, the Casimir energy densities for $z=1,3,5$ are given, respectively, by:
\begin{eqnarray}
    \mathcal{E}_{C}^{z=1}
    =
    - \frac{(qB)^{3/2}}{4\pi}
    - \frac{(qB)^{2}}{8\pi\,r^{1/2}} 
    \left( 1 + \ln \left( \frac{qB}{4} \right) \right) 
    - \frac{(qB)^{5/4}\,r^{1/4}}{4\pi^{2}}\,e^{-2\,\pi \sqrt{\frac{qB}{r}}} \,,
\end{eqnarray}
\begin{eqnarray}
    \mathcal{E}_{C}^{z=3}
    =
    - \frac{l^{2}\,(qB)^{5/2}}{4\pi}
    - \frac{l^{2}\,(qB)^{3}}{64\pi\,r^{1/2}} 
    \left( 7 + 6\ln \left( \frac{qB\,l^{4/3}}{4} \right) \right) 
    + \frac{3\,l^{2}\,(qB)^{7/4}\,r^{3/4}}{8\pi^{3}}\,e^{-2\,\pi \sqrt{\frac{qB}{r}}} \,,
    \nonumber\\
\end{eqnarray}
\begin{eqnarray}
    \mathcal{E}_{C}^{z=5}
    =
    - \frac{l^{4}\,(qB)^{7/2}}{4\pi}
    - \frac{l^{4}\,(qB)^{4}}{384\,\pi\,r^{1/2}} 
    \left( 37 + 30\,\ln \left( \frac{qB\,l^{8/5}}{4} \right) \right)
    - \frac{15l^{4}\,(qB)^{9/4}\,r^{5/4}}{16\pi^{4}}\,e^{-2\,\pi \sqrt{\frac{qB}{r}}}
     \,.
     \nonumber\\
\end{eqnarray}
As in the previous section, we perform a series expansion on the perturbation $r$ (see Eq. (\ref{rreduce})). The force expressions for each $z$ value are functions of integrations over the roughness function $\hat{f}$ (Eq. (\ref{M})), similarly to what occurs in the weak field limit (see Eqs. (\ref{forcedensityfinalz=1}) and (\ref{forcedensityfinal})). Then, as a last step, we use a periodic function for the roughness, in the same way as was done in the weak field limit: $ f(x) = \xi\cos(2\pi x/L)$. With this perturbative function the following condition arises: $qB\xi^{2}\gg 1$. This condition allows the integrations to be reduced, hence the Casimir force densities are given by
\begin{eqnarray}
    \mathcal{F}_{C}^{z=1}
    =
    \frac{(qB)^2}{8\pi^2}
    \left( 1 + \ln \left( \frac{qB}{4} \right) \right) 
    \left( 1 + \frac{3\xi^{2}}{4a^{2}} \right)
    - \frac {(qB)^{5/4}}{8\,\pi^{3/2}a^{3/2}}
    \left( 1 + 4a\sqrt{qB}\right) e^{-2\,a\sqrt{qB}}
    \,,
    \nonumber\\
\end{eqnarray}
\begin{eqnarray}
    \mathcal{F}_{C}^{z=3}
    =
    \frac{l^{2}(qB)^3}{64\pi^2}
    \left( 7 + 6\ln \left( \frac{qB\,l^{4/3}}{4} \right) \right) 
    \left( 1 + \frac{3\xi^{2}}{4a^{2}} \right)
    + \frac{3\,l^{2}(qB)^{7/4}}{16\,\pi^{3/2}a^{5/2}}
    \left( 3 + 4a\sqrt{qB}\right) e^{-2\,a\sqrt{qB}}
    \,,
    \nonumber\\
\end{eqnarray}
\begin{eqnarray}
    \mathcal{F}_{C}^{z=5}
    =
    \frac{l^{4}(qB)^4}{384\pi^2} 
    \left( 37 + 30\ln \left( \frac{qB\,l^{8/5}}{4} \right) \right) 
    \left( 1 + \frac{3\xi^{2}}{4a^{2}} \right)
    - \frac {15\,l^{4}(qB)^{9/4}}{32\,\pi^{3/2}a^{7/2}}
    \left( 5 + 4a\sqrt{qB}\right) e^{-2\,a\sqrt{qB}}
    \,.
    \nonumber\\
\end{eqnarray}
As in previous findings, when $z$ takes even values, the force density vanishes. For $z=1$ we can note that the force density contains a term independent of the separation distance $a$ (this term contributes because we have considered the absence of the magnetic field on the outside of the plates), and another term that depends on the roughness factor of order two that must be considered. For $z=3,5$ we see that the forces have a structure similar to the case $z=1$, but it has a strong dependence on the scaling factor $l$. This scaling factor plays a crucial role in determining the orientation of the force, due to its presence in the logarithm function. As the plate separation \( a \) increases, the Casimir force density approaches the global factor that only depends on the magnetic field for a fixed $l$, while the exponential contribution becomes negligible. As anisotropy increases, the magnitude of the force decreases and its orientation changes (see Fig.~\ref{fig2}(a)). In Fig.~\ref{fig2}(b), for a fixed \( a \), increasing the value of \( qB \) leads to a rapid decrease in force density, particularly for larger anisotropic factors \( z \). The force density for \( z=1 \) is positive and several orders of magnitude greater than for \( z > 1 \).

\begin{figure}[H]
\begin{minipage}[b]{0.5\linewidth}
    \centering
    \includegraphics[width=1.0\linewidth]{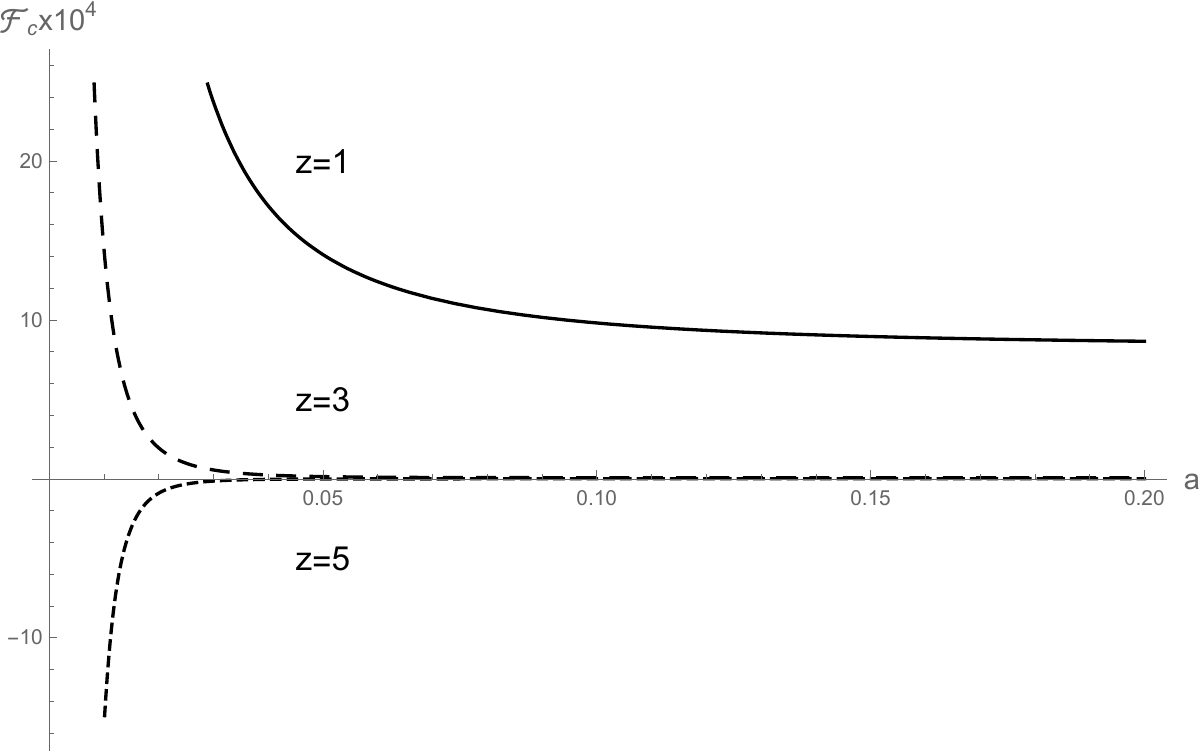} 
    \caption*{(a)}
    \label{fig2a}
    \vspace{4ex}
\end{minipage}
\begin{minipage}[b]{0.5\linewidth}
    \centering
    \includegraphics[width=1.0\linewidth]{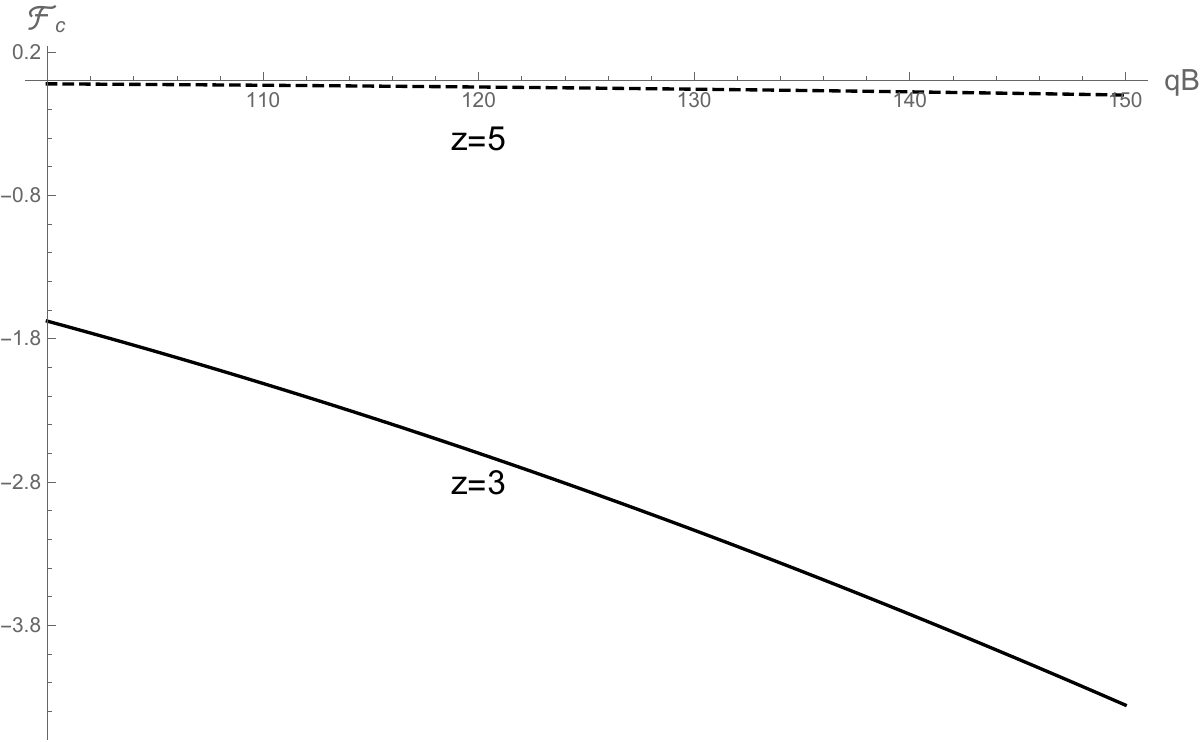}
    \caption*{(b)}
    \label{fig2b}
    \vspace{4ex}
\end{minipage}
\caption{Casimir force density as a function of separation distance and magnetic field, for anisotropic scaling parameters $z$. (a) The solid, large dashed, and dashed curves represent roughness ($\xi=0.05$ (units of length)) with a constant uniform magnetic field ($qB=1000$ (units of length$^{-2}$)). (b) The two curves show the Casimir force density for fixed separation distance ($a=0.5$ (units of length)) and roughness ($\xi=0.05$ (units of length)).}
\label{fig2}
\end{figure}


\section{Conclusions}

We investigate the case of uncharged, isolated, rough parallel plates embedded in a $3+1$-dimensional manifold. Vacuum fluctuations are induced by an anisotropic charged-scalar quantum field with Dirichlet boundary conditions. The Lagrangian density is formulated within the framework of a Ho\v rava-Lifshitz-like theory. Additionally, we analyze the effect of a constant and uniform magnetic field inside the parallel plates and oriented perpendicular to them. To address the roughness of the plates, we perform a change of variables that flattens their borders, treating the remaining terms as a perturbative roughness incorporated into a potential. To determine the eigenvalues, we apply perturbation theory up to first order and utilize the $\zeta$-function regularization method. We present an explicit case where the perturbation of parallel plates exhibits periodic behavior. As in previous works, we find that the Casimir energy and force vanish when the anisotropic scaling factor takes even values, independently of any other physical factors in the result \cite{Borquez:2023cuf,Borquez:2023ajx}. Contributions from roughness up to the second order in perturbations, as well as the magnetic field, are present as expected. For odd values of the anisotropic factor, we examine the limiting cases of the magnetic field. In the weak-field case, magnetic effects are minimal. Consequently, we find that energy and force densities are significantly modified by roughness. As found in Eq. (\ref{FuerzaZ}), unlike in the case of $z = 1$, the force density depends on the scaling factor $l$ for any value of the anisotropic factor. As higher values of $z$ are chosen, the force density strongly depends on $l$ decreasing in magnitude. The anisotropic scaling factor plays a crucial role in determining the direction of the force. On the other hand, when taking into account the roughness we notice that in each case the force density increases in magnitude compared to the planar case.

In the strong-field regime, for $z>1$ we see that the forces have a structure similar to the case $z = 1$, but it has a strong dependence on the scaling factor $l$. As the plate separation $a$ increases, the Casimir force density approaches the global factor that only depends on the magnetic field for a fixed $l$. Anisotropy directly affects the orientation of the force density. Besides, the increasing the value of $qB$ leads to a rapid decrease in the force density, particularly for larger anisotropic factors $z$, however, the force density for $z = 1$ is positive and more dominant than the $z > 1$ cases.

Since the Casimir effect can be measured with precision, the impact of roughness could contribute to estimating the Lorentz-breaking parameters in the theory. Therefore, the Casimir effect is essential for verifying the breaking at high energies. By adding other factors to the result, such as temperature and different boundary conditions, is expected to modify the Casimir effect with rough edges. By defining the perturbation as a periodic function significantly simplifies the final result. If more complicated functions are chosen for the perturbation, other resolution methods must be used. We will consider these cases in future work.


\end{document}